 \documentclass[aps,prb,twocolumn,amsmath,amssymb,
   superscriptaddress,floatfix,showpacs]{revtex4}

\usepackage{epsfig}
\usepackage{graphicx}
\usepackage{color}

\newcommand{\Zn}{ZnCu$_3$(OH)$_6$Cl$_2$}
\newcommand{\Cu}{Cu$_3$V$_2$O$_7$(OH)$_2\cdot$2H$_2$O}

\newcommand{\bea}{\begin{eqnarray}}
\newcommand{\eea}{\end{eqnarray}}
\newcommand{\beq}{\begin{equation}}
\newcommand{\eeq}{\end{equation}}

\def \be{\begin{equation}}
\def \ee{\end{equation}}
\def \ba{\begin{array}}
\def \ea{\end{array}}
\def \bea{\begin{eqnarray}}
\def \eea{\end{eqnarray}}

\def \etal{{\it {et al}}}

\def\kago{kagom\'{e}}
  \def\HAKAF{\mathcal{H}_{\rm AHM}}

\def\S{{\mathbf{S}}}

\def \journal#1#2#3#4{{#1\ {\bf #2}, #3 (#4)}} 
\def \sec#1{{Sec.~\ref{#1}}}
\def \fig#1{{Fig.~\ref{#1}}}
\def \eqn#1{{Eqn.~(\ref{#1})}}

\begin{document}

\title{ Thermodynamic properties of the Spin-1/2 
Heisenberg Antiferromagnet with Anisotropic Exchange
on the Kagom\'{e} Lattice: Comparison with Volborthite 
}

\begin{abstract}
Thermodynamic properties such as 
magnetic susceptibility and specific heat
have been computed for the Heisenberg Antiferromagnet with 
spatially anisotropic exchange on the kagom\'{e} lattice
on clusters up to $N=24$ spins 
from the full spectra obtained by exact diagonalization. 
This  approach is shown to provide a  
good represention of these thermodynamic properties
above temperatures of about $J_{\rm av}/5$ where
$J_{\rm av}$ is an average of the coupling constants.
Comparison with experimental Volborthite data 
obtained by Hiroi {\it {et al}}
[J. Phys. Soc. Jpn. {\bf 70},3377 (2001)]
shows that Volborthite  is best described
by a model  with nearly isotropic exchanges
in spite of the significant distortion of the 
kagom\'{e} lattice of magnetic sites in this compound 
and suggests that additional interactions are present.
Comparison of the specific heat at low temperature
raise the possibility that the density of states at low energy
in Volborthite might be much lower than in the Heisenberg model.
Magnetization curves under an applied field  of the model
are also investigated. The $M=1/3$ plateau is
found to subsist in the anisotropic case 
and extend to lower field with  increased anisotropy.
For sufficient anisotropy, this  plateau would 
then be observable 
for a field reasonably accessible to experiment.
The absence of a plateau well below $\sim$ 70 Teslas
would further support a nearly isotropic model. 
\end{abstract}

\author{Philippe Sindzingre}
\email{phsi@lptmc.jussieu.fr}
\affiliation{Laboratoire de Physique Th\'eorique de la Mati\`ere
condens\'ee, Univ. P. et M. Curie, 75252 Paris Cedex, France}
%\date{July 4th, 2007}
\date{\today}

\maketitle

\section{Introduction}
\label{sec:introduction}

Frustrated quantum magnetic insulators have received a large attention 
since many years, due to the possibility of observing unconventional
behavior~\cite{ml2005}. 
Amongst these, antiferromagnets with a {\kago} lattice
of spin-1/2 appear to be promising candidates.
However, despite numerous theoretical investigations, the nature
of the ground-state of such spin systems remain an open question.
In the most studied case of the spin-1/2 Heisenberg model,
numerical studies have concluded to the absence of
long-range magnetic (N\'eel) order
and the presence of an unusally large density of states at low 
energy~\cite{elser89,ze90,ce92,sh92,le93,lecheminant97,web98,smlbpwe00}.
But many questions such as the existence of spontaneously broken symmetries
or even the existence of a finite gap to magnetic excitations 
are not settled~\cite{ms_07,sl_07}.

In recent years, two promissing experimental realization 
of a spin-1/2 {\kago} antiferromagnet have been synthezied and studied
for their magnetic properties: 
\Cu ({\it Volborthite}) ~\cite{hiroi_2001,bert_2005}
and recently \Zn ({\it Herbertsmithite})~\cite{
helton06,mendels06,ofer06,imai07,ms2007,devries07,SHlee97}.
Both compounds do not show any signs of ordering 
down to temperatures well below the exchange coupling strength
and could be amongst the first 
realizations of a 2D quantum spin liquid~\cite{anderson73,ml2005}.
Yet they appear to deviate somewhat from a perfect realization
of the {\kago} Heisenberg model. They differ in different ways.
Herbertsmithite has a perfect {\kago} geometry  but
contains a signicant (probably intrinsic) percentage of impurity spins 
arising from antisite disorder~\cite{ms2007,devries07,SHlee97} 
and may also deviates from the Heisenberg model due to the 
presence of additional interactions,
possibly Dzyaloshinskii-Moriya interactions~\cite{rs07,rs07b}.
In particular, the effect of the impurity spins 
which have significant interactions between themselves and
with the spins residing on the {\kago} sites seems to be
not easy to modelize.
This complicates the analysis of the experimental data 
and the determination of the relevant model describing Herbertsmithite.
By contrast, Volborthite has the advantage of being available with
a very low impurity content.
However, it presents a deformed {\kago} geometry.
There, the equilateral {\kago} triangles are distorted into
isoceles triangles which suggest that two of the
nearest neighbor exchange constants are different from the third.
But to which extent the exchange couplings differ and 
whether  an  Heisenberg model with such a spatial anisotropy
of exchange couplings is sufficient to describe Volborthite
remained an open question.

\begin{figure}% [h]
  \includegraphics[width=6cm]{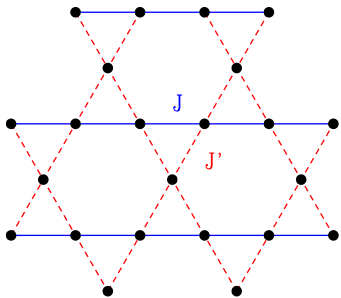}
\caption{(color online)
Anisotropic {\kago} model.
The {\kago} sites are show as black dots.
$J$ is the coupling constant between nearest neighbour
spins on horizontal lines
(solid lines in blue)
whereas $J'$ is the coupling constant between a middle spin
and its nearest neighbour spins (dashed red lines).
}
\label{aniskago}
\end{figure}

The main purpose of this paper is to study this question by
comparing the thermodynamic quantities measured  by Hiroi {\etal}~\cite{hiroi_2001}
with the results obtained from exact diagonalization (ED)
of an anisotropic Heisenberg model (AHM), described by the Hamiltonian:
\be
\HAKAF = J \sum_{[i,j]} \S_i \S_j + J' \sum_{\langle k,i \rangle}
 \S_k \S_i\;.
\label{hamilt_akago}
\ee
where the symbols $[i,j]$ note pair of nearest neighbour sites 
on the horizontal chains with exchange coupling $J$
and $\langle k,i \rangle$  pair of nearest neighbour sites
between the middle sites and sites on the chains
with exchange coupling $J'$ (see~\fig{aniskago}).
$J$ and $J'$ are taken antiferromagnetic (positive).
The ratio $\alpha=J/J'$ 
will be used to measure the anisotropy of the couplings
and the average $J_{\rm av}= (J + 2 J')/3$  to set the energy scale.
The classical AHM has a ferrimagnetic ground-state for 
$\alpha\le 0.5$~\cite{aye_07,yae_07} which subsists for the spin-1/2
model  over nearly the same range of values $\alpha$.
We shall thus limit ourselves to the region $\alpha\gtrsim 0.5$, where the 
ground state has zero magnetization.
Thermodynamic properties were computed from the full spectra
of clusters up to $N=24$ spins which is shown to yield  results that 
may be considered as reliable estimate of these quantities
in the thermodynamic limit down to temperatures $T\sim 0.2 J_{\rm av}$ or
even below.

We focus on the comparison of the experimental and numerical
magnetic susceptibility $\chi$,
acurately measured  by Hiroi {\etal}~\cite{hiroi_2001}
over a large range of temperatures $T$,
since it is presently the best known of the thermodynamic quantities.
As discussed below, the comparison indicates that Volborthite
is best described by the Heisenberg model with at most a small anisotropy 
and reveals that additional interactions are present in Volborthite
besides the nearest neighbour couplings.

Besides the magnetic susceptibility, this paper reports
the results obtained from these ED calculation for the 
magnetic heat capacity $C_v(T)$ and the magnetization curves $M(H)$
under an applied magnetic field $H$.
The available data for Volborthite are however still limited,
which hinder the use of this quantities for the determination
of the model which could be relevant to describe Volborthite.
The total heat capacity was measured
by Hiroi {\etal}~\cite{hiroi_2001} but the magnetic part
is somewhat uncertain, especially at high temperatures,
since the lattice part is poorly known. 
Yet, as discussed below, the comparison at low temperature
raises the possibility that the density of states at low energy
in Volborthite might be much lower than in the Heisenberg model.
The magnetization of Volborthite was only measured by Hiroi \etal~\cite{hiroi_2001}
under low fields (up to a few Tesla) which only  gives access to the very low part
of the magnetization curve.
As shown below, the  anisotropic Heisenberg model show a
plateau at one third of the saturation of the magnetization
which starts at a field that decrease with increasing anisotropy.
Its location (or absence) will provide further information 
on the nature of the model which is relevant to Volborthite.

This paper does not  address the question of the nature
of the ground state of the AHM which has been studied
recently by Yavors'kii, Apel and Everts~\cite{aye_07,yae_07} 
or Wang\etal~\cite{wvk_07} using mean field approaches.
The analysis of the low energy spectrum of the AHM 
and this question will be the subject of a forthcoming manuscript.
ED results for the  magnetic susceptibility and
the heat capacity  have been reported by Wang{\etal}~\cite{wvk_07}
but where not compared to the experimental data and where obtained
only for a  cluster of $N=12$ spins.
Wether this very small size systems could be sufficient for the comparison
with experimental data was unclear.
%which could have been  too small to allow a firm estimation of these thermodynamic
%quantities  over a large range of temperatures in the limit $N\rightarrow\infty$.
%which the present results, obtained with up to $N=24$ spins, enable. 

The paper is organized as follows. Numerical results
obtained from the AHM model for the magnetic susceptibility
and the heat capacity are compared with experiment in \sec{sec:chi}
and \sec{sec:cv}, repectively.
The magnetization curves of the AHM model and the possibility
of the detection of a magnetization plateau experimentally
are considered in \sec{sec:magnetization}.
\sec{sec:summary} summarizes the  results.  

\begin{figure}%[h]% [tb]
 %\centering
 \includegraphics[width=7.5cm]{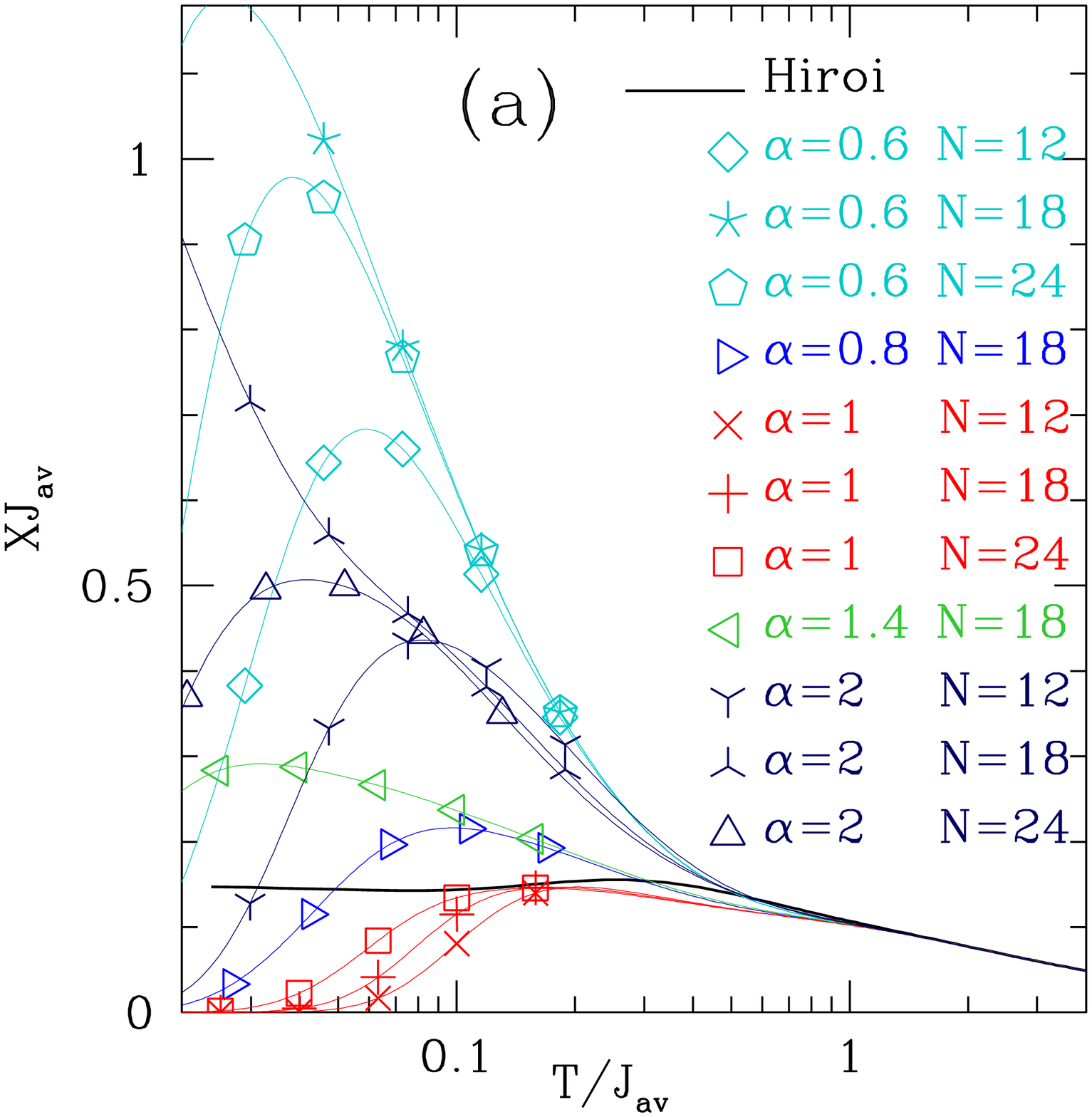}

 \includegraphics[width=7.5cm]{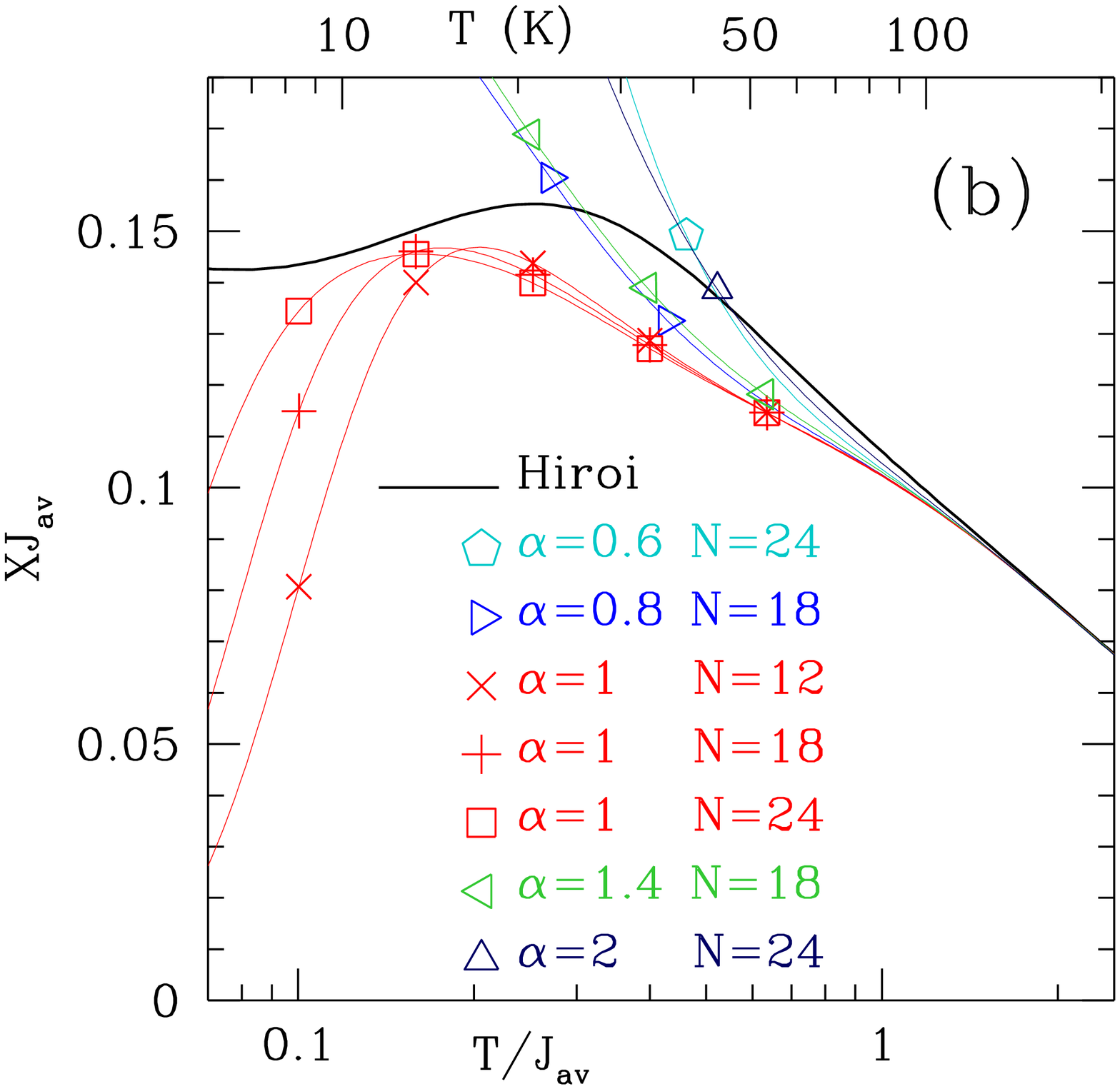}
\caption{(color online)
Magnetic susceptibility $\chi$ vs temperature $T$ scaled
to $J_{\rm av}=(J+2J')/3$ (see \eqn{hamilt_akago}).
The curves obtained from exact diagonalization (ED)
for different values of the anisotropy $\alpha=J/J'$
and number of spins $N$  are shown by light coloured lines,
specified by different symbols, as indicated in the legende.
The black heavy line connects
the experimental data of Hiroi {\etal}~\cite{hiroi_2001}
with $T$ scaled to $J_{\rm av}=84.1 K$.
(a) includes ED results  for $0.6\le\alpha\le 2$
and cluster sizes $N=12,18,24$.
(b) displays an enlarged region  of (a) for a better comparison
of the ED data at small anisotropy with experimental data
(only part of the ED data at large anisotropy are shown for clarity).
\label{fig_chi}}
\end{figure}

\section{Magnetic susceptibility}
\label{sec:chi}
ED results for the magnetic susceptibility $\chi(T)$
for different cluster sizes and selected values of the anisotropy $J$
are plotted in \fig{fig_chi} to gether with the experimental data 
scaled with $J_{\rm av}=84.1 K$. 
This scaling  was found by Hiroi {\etal} to 
enable a fit of the data to the high temperature series expansion  
of the isotropic model for $T\gtrsim 2.5$.

A comparison of the $\chi(T)$ obtained for $J=0.6, 1, 2$
at the  different sizes  indicate that
the ED results become to be well converged to their 
thermodynamic values down to temperature below $T/J_{\rm av}\sim 0.2$
once $N\ge 18$ whereas the convergence of the $N=12$ results is
slightly poorer.
We can thus safely compare the $N=18$ numerical data 
for $0.6 \le J \le 2$ with experiment over this range of temperature.
\footnote{%%
The calculation of the full spectrum for $N=18$ is very rapid and 
provide results that are converged down to a temperature
below those atained from high temperature expansions 
or even improved similar methods.
This suggest an alternative  approach to derive high temperature expansions
which would consist to fit the coefficients of the expansion
to the ED results over the range of temperatures where the ED
results may be considered as converged.
}%% end footnote
Having assessed the convergence of the ED results, we 
examine  the effect of the anisotropy of $\chi$
and compare with experiment.

In \fig{fig_chi}~(a) one sees that the ED $\chi(T)$ are unsensitive
to the anisotropy  for $T/J_{\rm av}\gtrsim 1$ 
--and coincide very well with the experimental data.
However, anisotropy has a large effect on $\chi(T)$ at lower T.
Between $T/J_{\rm av}\sim 1$ and $T/J_{\rm av}\sim 0.1-0.2$,
where the ED results are subjected to little size effects
and can be trusted as converged, one observes a strong increase
of $\chi$ with anisotropy which 
for $\alpha=0.6$ or $2$ reach several times  its value
for $\alpha=1$ --and the experimental value.
\footnote{%%
Note that in this range of temperature two different values of $\alpha$,
one $>1$  and the other $<1$ lead to rather similar $\chi(T)$.
It is uneasy to differentiate between the two
case on the sole basis of the magnetic susceptibility.
}%% end footnote
Qualitatively, this increase of the susceptibility 
can be understood as follows:
(i) as $\alpha$ decreases toward $0.5$, the system approachs
the ferrimagnetic phase where $\chi$ diverges at $T=0$,
(ii) for large $\alpha$ the middle spins become
quasi disconnected from the chain spins and from one another,
behaving as quasi free spins under an applied field
(until the field fully polarize these middle spins
leading to the formation of a plateau at one third of the 
saturated magnetization-see \sec{sec:magnetization}).
As shown in \fig{fig_chi}~(a) such a large increase
of $\chi$ at low $T$ does not occur in Volborthite.
This leads to conclude that a AHM with a large anisotropy
is not appropriate to describe Volborthite.
However, as can be seen in \fig{fig_chi}~(b),
a AHM  with $\alpha\sim 1$ is neither fully satisfactory.
The shape of the ED $\chi(T)$ curves differ somewhat from
the  experimental one.
The AHM displays a bump in $\chi(T)$ at $T/J_{\rm av}\sim 1$
absent in Volborthite.
For small or moderate anisotropy the ED $\chi(T)$ are 
slightly smaller than the  experimental $\chi(T)$ in the range 
$0.2\lesssim T/J_{\rm av}\lesssim 1$.
In the AHM, the maximum of $\chi(T)$ always occurs
at a lower temperature than in Volborthite.
This reveals the presence in Volborthite of additional
interactions  between the spins besides those included 
in the AHM.

\begin{figure}%[tb]
  \includegraphics[width=7.5cm]{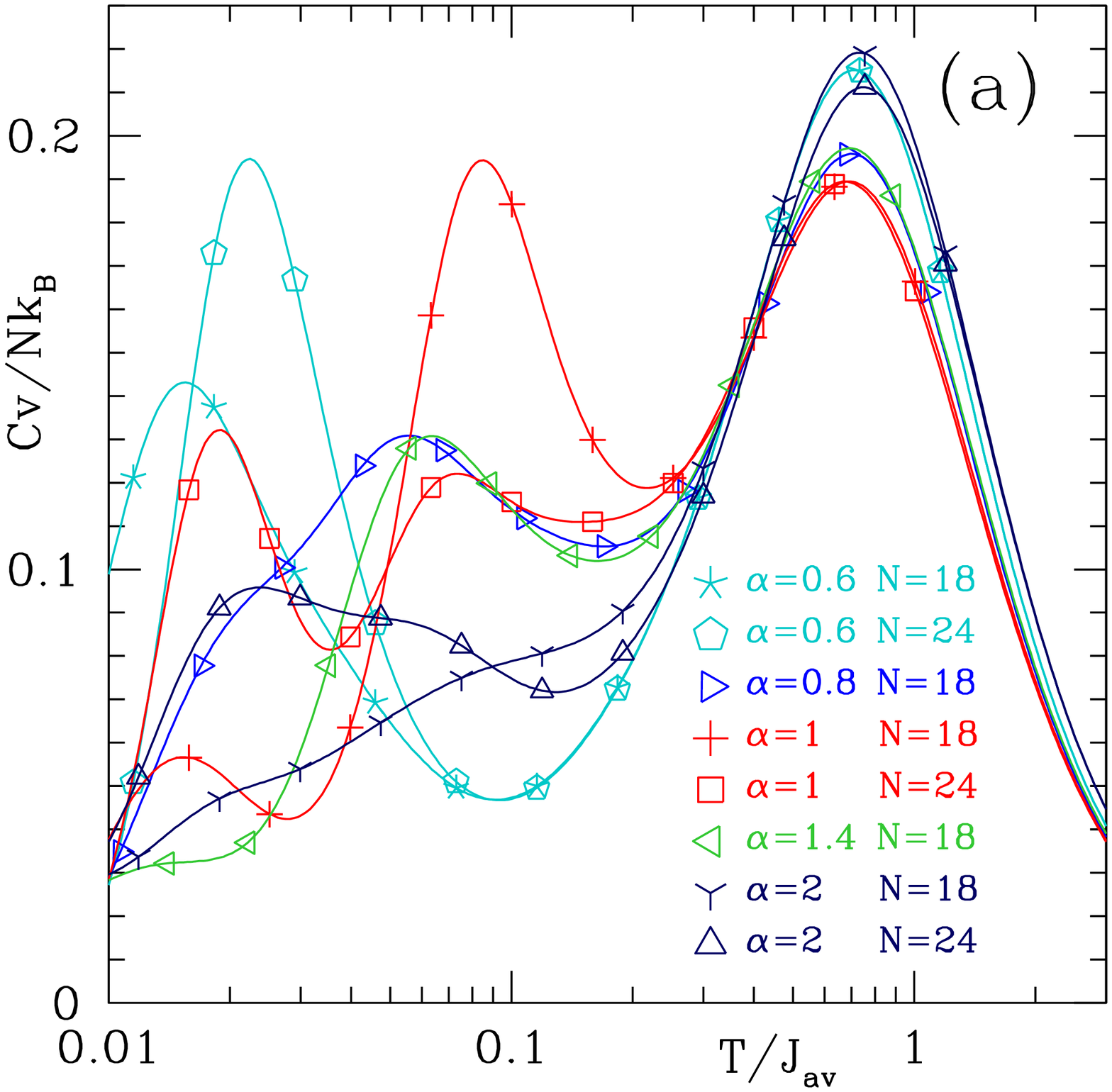}

  \includegraphics[width=7.5cm]{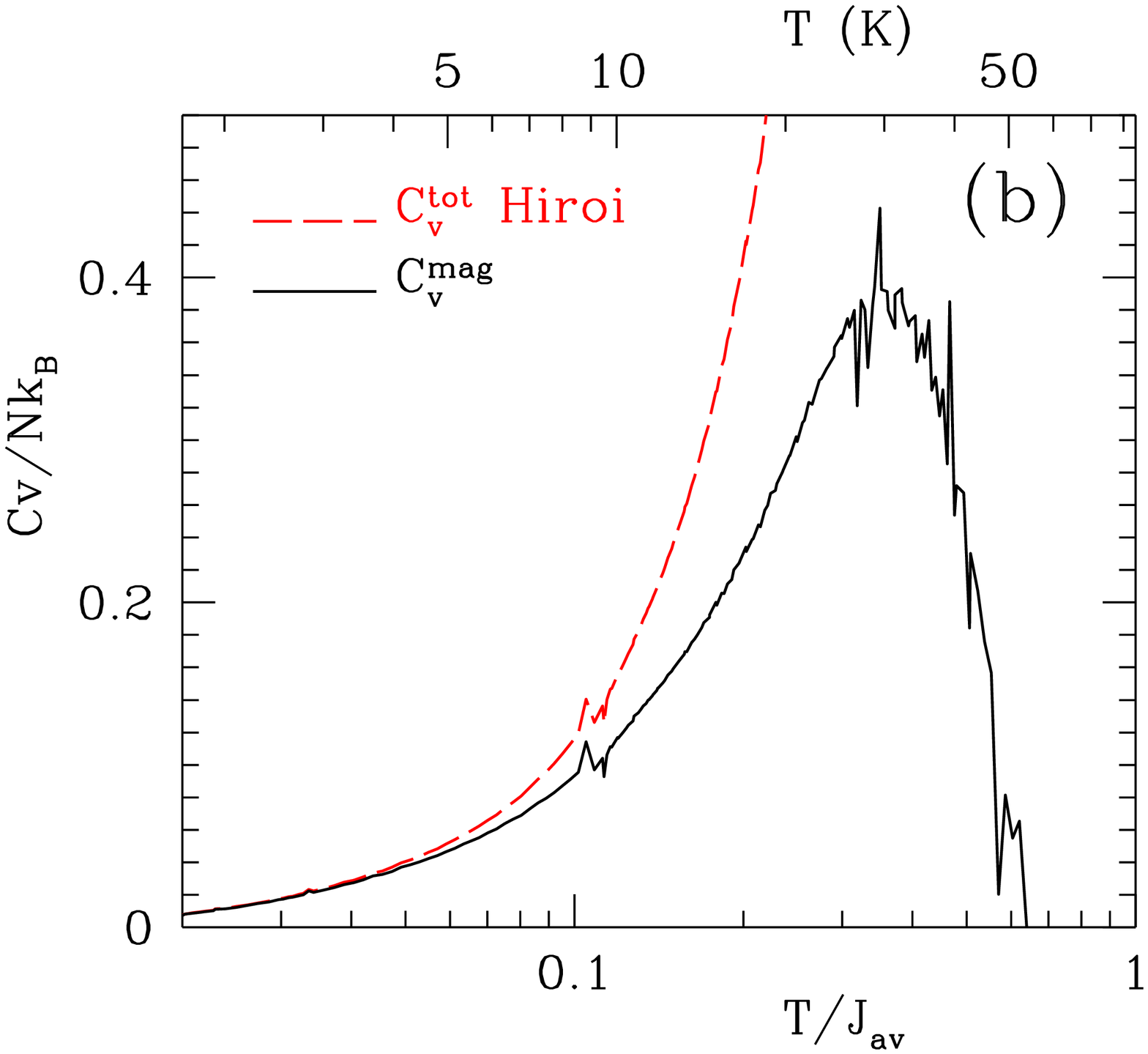}
 \caption{(color online)
Specific heat $C_v$ vs temperature $T$.
(a): results of exact diagonalizations (ED)
--see caption of \fig{fig_chi}--.
(b): experimental results of Hiroi {\etal}~\cite{hiroi_2001}
with $T$ scaled to  $J_{\rm av}=84.1 K$.
$C_{v}^{tot}$ is the total specific heat measured.
$C_{v}^{mag}$ is the magnetic part derived  by Hiroi {\etal}
by substracting from $C_{v}^{tot}$ the lattice contribution
estimated from a Debye model.
\label{fig_cv}}
\end{figure}

\begin{figure}%[tb]
  \includegraphics[width=7.0cm]{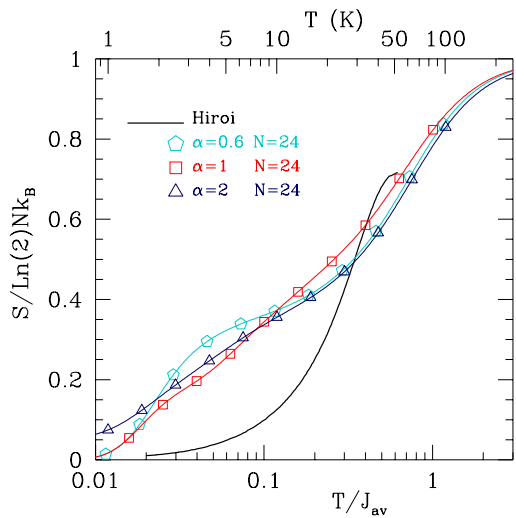}
 \caption{(color online)
Integrated entropy  $S$ vs temperature $T$.
\label{fig_entropie}}
\end{figure}

\section{Specific heat}
\label{sec:cv}
ED results for the specific heat $C_v(T)$
are displayed in \fig{fig_cv} (a) 
for the same values of the anisotropy $\alpha$ as in \fig{fig_chi}.
For the sake of clarity experimental data are shown below
in \fig{fig_cv} (b).
The ED and experimental results for the
integrated entropy $S(T)=\int_0^T c_v(x)/xdx$
are plotted in \fig{fig_entropie}.

The ED results for $C_v$ show little size effects
and can be considered as (quasi)converged  for $T/J_{\rm av}\gtrsim 0.2$
like those obtained for the susceptibility $\chi$.
In this range of temperature $C_v$ exhibit a broad peak
around $T/J_{\rm av}\sim0.6-0.8$ which increases and shift
to higher $T$ with increased anisotropy but only very slightly. 
This is the usual main peak, associated to large energy excitations,
rather insensitive to the detail of the interactions in the AHM.
However, the integrated entropy under this peak 
$\sim S(T=\infty) - S(T=0.2J_{\rm av})$
(see \fig{fig_entropie}) is only about 1/2 of the total entropy.
Thus, as already found in the isotropic case~\cite{smlbpwe00} 
an  unusually large part of the entropy is located at low 
temperature, which reveals a large density of states at low energy.
For all values of the anisotropy, such a feature is seen to remain.

The exact location of this low temperature entropy 
--i.e. wether there exist one (or more) extra peak at low $T$--
is however subject to some uncertainties. 
The size effects on $C_v$ at low $T$  are large.
This make the extrapolation of $C_v$ in the thermodynamic limit difficult.
In the case of the isotropic model, the finite size $C_v$, 
estimated for clusters up to $N=36$, were found to display an extra peak 
at about $T/J_{\rm av}\sim 0.1$  or below~\cite{smlbpwe00},
associated with a very large number of states,
which is likely to survive in the thermodynamic limit~\cite{mb05}
(and a small peak at very low $T$ $\lesssim 0.02 J_{\rm av}$,
associated to a few states, 
which may disappear in the in the thermodynamic limit).
The exact location of this extra peak in the thermodynamic limit
is not known precisely, but its presence is necessary to account 
for the large entropy at low $T$, unless $C_v$ increases unsually fast with $T$
for $T\rightarrow 0$, e.g. as $T^{\beta}$ with $\beta<1$~\cite{mb05,ms2007}.
For the same reasons a low $T$ peak  seems likely in the anisotropic case.
The finite size results in \fig{fig_cv} (a) indicate that
this peak shifts to lower T with increasing anisotropy.

The experimental data of Hiroi {\etal}~\cite{hiroi_2001},
are plotted in \fig{fig_cv} (b)  where we show 
both the total heat capacity $C_{v}^{tot}$ measured and 
the estimated magnetic part $C_{v}^{mag}$.
$C_{v}^{mag}$ was derived, in absence of a non magnetic isomorph, 
by substracting  the lattice contribution
estimated from a Debye model with a temperature of $\theta_{D}=320 K$.
As pointed out by Hiroi {\etal}  this crude approximation of the 
lattice contribution affect the accuracy of $C_{v}^{mag}$, 
especially at high temperature.
The resulting $C_{v}^{mag}$ vanishes around $T\sim 60 K$.
By contrast, as seen in \fig{fig_chi}, 
the experimental magnetic susceptibility
is still large at $T\sim 60 K$.
Moreover, the integrated entropy up to $T=60 K$  is only $\sim70\%$
of the total magnetic entropy (see \fig{fig_entropie}).

The comparison between $C_{v}^{mag}$ which is only accurate at low $T$
and the  ED $C_v$ which are only converged at high $T$ is not easy. 
$C_{v}^{mag}$ is quite different from the ED $C_v$ of \fig{fig_cv} (a).
$C_{v}^{mag}$ exhibits a peak at $T\sim 30 K$.
The height of this peak is about twice  larger
than the height of the main peak of the ED results.
If the temperature of the experiment are scaled to $J_{\rm av}=84.1 K$,
this peak is located at a temperature $T/J_{\rm av}\sim 0.3$ 
This is about half of the temperature of the main peak of the ED $C_{v}$.
The differences for $T/J_{\rm av}\gtrsim 0.2$ are thus quite large.
But $C_{v}^{mag}$ is rather uncertain in this range of temperature.
Most likely, the peak of the real $C_{v}^{mag}$ is located at a
temperature higher than $\sim 30 K$.
Yet one may notice that the integrated entropy up to to $60 K$ (see \fig{fig_entropie})
is similar to the one found for the AHM up to $T/J_{\rm av}\sim 0.7$.

At lower $T$, however, the lattice contribution is small and  $C_{v}^{mag}$
more accurate. But there also, one sees quite distinct behavior.
$C_{v}^{mag}$ increases much less rapidly with $T$.
There is no low $T$ peak.
As a result the shapes of the integrated entropy widely differs
(see~\fig{fig_entropie}).
The integrated entropy obtained from $C_{v}^{mag}$
is much smaller than in the AHM which
reveals that the density of states at low energy 
could be  much smaller in Volborthite than in the AHM model.
This may further suggest that some modification of the AHM
are needed to describe Volborthite. 

\begin{figure}%[tb]
  \includegraphics[width=8.0cm]{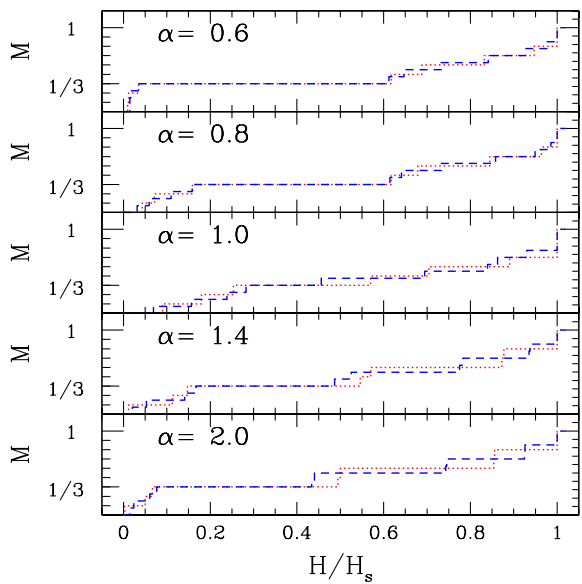}
 \caption{(color online)
Magnetization curves for different value of the anisotropy $\alpha$,
obtained from exact diagonalization on clusters of $N=18$
(dotted red lines) and $N=24$ (dashed blue lines) spins. 
% $m_s$ is the magnetization  at saturation and 
The magnetization $M$ is normalized to 1.
$H_s$ the saturation field.
\label{fig_magnetization}}
\end{figure}

\section{Magnetization curves}
\label{sec:magnetization}
The magnetization of the isotropic model $\alpha=1$
has already been the subject of many studies. 
The model displays a plateau at $M=m/m_s=1/3$ of the saturated  magnetization $m_s$,
which was interpreted as valence bond state with 
a $\sqrt{3}\times\sqrt{3}$ superstructure~\cite{cabra_05}
and a jump of height $\delta M=2/9$ to saturation
%at a field $H_s=3J$ 
which arises degenerate localized magnons~\cite{richter_05}.
The $M=1/3$ plateau was aslo reported  to subsist 
in the anisotropic case~\cite{cabra_03}.
But this case remained much less studied.

The magnetization curves obtained from the ED spectra 
are ploted in \fig{fig_magnetization} 
for different value of the anisotropy $\alpha$ 
in function of $H/H_s$ where $H_s$ is the saturation field.
The  $M=1/3$ plateau %can be recognized. 
subsist in the anisotropic case.
The magnetization jump close to saturation disappear for $\alpha<1$
and is replaced for $\alpha>1$ by a smaller jump 
(proportional to the number of horizontal lines in the cluster)
which will disappear in the thermodynamic limit.
The spin structure in the $M=1/3$ plateau corresponds, 
for $\alpha < 1$, to the ferrimagnetic phase
(present in zero field for $\alpha\lesssim 0.5$),
whereas for $\alpha > 1$ it corresponds to full 
polarization of the middle spins.
One may notice that the width of the $M=1/3$ plateau increases 
and the lower field $H_1$ for which it appears decreases 
with increased anisotropy.
For a large anisotropy this field become quite low.
The value of $H_s$ is $3J'=3J_{\rm av}/(2+\alpha)$ if $\alpha<1$
and $2(J'+2J)=3J_{\rm av}(1+2\alpha)/(2+\alpha)$ if $\alpha>1$ 
(with $H_s=3J_{\rm av}=3J$ if $\alpha=1$).
\fig{fig_magnetization} indicate that $H_1\sim 0.3 H_s$ if $\alpha=1$ 
but becomes  $\lesssim 0.1 H_s$ for $\alpha\lesssim 0.7$
or  $\alpha\gtrsim 2$. 
If $J_{\rm av}=84 K$, $H_s\sim 200$ Teslas, so $H_1\sim 70$ Teslas
if $\alpha\sim 1$, but
for a large anisotropy $H_1$ could  be  $\sim 20$ Teslas
i.e. in a range accessible to experiments.
If a compound  would be well described by the AHM Hamiltonian,
the presence of a large anisotropy would be signaled by
an observable  magnetization plateau at low field, 
to gether with a large peak in the magnetic susceptibility 
at low temperature(as seen in \sec{sec:chi}).
Additional interactions to the AHM Hamiltonian may 
have different effects on the location of the $M=1/3$ plateau
and the susceptibility. 
But most likely a magnetization plateau at low field
would imply a large peak in the magnetic susceptibility at low temperature.
Since the latter feature is not found  in Volborthite,
it is probable that the $M=1/3$ plateau (if any) is not
located at low field but around $\sim 70$ Teslas.
Presently, the magnetization has been only measured for fields up
to 7 Teslas~\cite{hiroi_2001} (and  no plateau has been seen).
Measurements at higher fields may be worth.
The absence of a plateau up to $\sim 70$ Teslas
would confirm a small anisotropy.
If a plateau is observed below, its location would be a valuable
information for the determination of the model 
appropriate to Volborthite.

\section{Summary}
\label{sec:summary}
In this work, we have investigated the thermodynamic properties
of the spin-1/2 spatially anisotropic Heisenberg model 
on the {\kago} lattice by means of exact diagonalization
in order to compare with available experimental data 
for Volborthite.

The exact diagonalization results for the magnetic 
susceptibility and specific heat are  found to
be well converged  to their values in the thermodynamic limit
down to temperatures $\sim 0.2 J_{\rm av}$ (and perhaps below)
for clusters of $N=24$ spins. 
The range  of temperature is significantly larger 
than the one accessible with present 
high temperature series expansion calculation~\cite{ey94,mb05}
and also somewhat larger than with other methods such as the
linked cluster expansion of Ref.\cite{rs07,rs07b}.

The comparison of the computed magnetic susceptibility
with the experimental data show that Volborthite is best
described by an Heisenberg model with at most a weak anisotropy.
This suggests that the spatial anisotropy of the couplings in Volborthite 
is weak in spite of the significant distortion  of the {\kago} lattice
in this compound. 
It also reveals the necessity to introduce  in the present model
other interactions in addition to the nearest neighbor exchanges
to allow a good  fit to the magnetic susceptibility of Volborthite
below $\sim J_{\rm av}$.

The comparison of the numerical specific heat with
experimental data reveal quite large differences
but is uneasy since the latter are rather
uncertain, especially at high temperature.
At low temperature, the integrated entropy of Volborthite
appears to be much smaller than in the model.
This suggests that the high density of states at low energy,
which is a carateristic of the Heisenberg model
may not be present in Volborthite.
Although, more accurate experimental data  will be
necessary to check this assumption, this may indicate
some significant differences between the physics of the 
Heisenberg model and Volborthite at low energy.
In Herbertsmithite, where the magnetic specific heat is also uncertain,
a similar comparison of the integrated entropy at low temperature
leads to an analogous conclusion~\cite{ms2007}.
For both compounds, this point seems worth further experimental
investigation and will put constraint on the additional
interactions which are to be introduced in the Heisenberg model
in order to better describe these compounds.

An other carateristic property of the {\kago} Heisenberg model
is the presence of a magnetization plateau at one third of the 
saturated magnetization.
Because of the large value of the average exchange in Volborthite
($\sim 84 K$), the saturation field may be larger than 200 Teslas
and the experimental study of the magnetization is uneasy.
For weak anisotropy the plateau may appears around $\sim 70$ Teslas
and its width may not be large.
Yet, the  width of the plateau is found to increases with anisotropy and
for large anisotropy  this plateau  starts at an  applied
field which may reach very small value more accessible 
to experiment. 
In view of the experimental data for the susceptibility,
the detection of a plateau at low field seems however unlikely.
Presently, the magnetization has been measured for fields up 
to 7 Teslas~\cite{hiroi_2001} and  no plateau has been seen.
The absence (or the presence) of a plateau up to much higher fields  
would provide further inside into the model that can describe Volborthite.

Although, it may not allow a perfect description of Volborthite,
the anisotropic Heisenberg model provide a starting point
and a reference for the understanding of this compound
(or compounds with a distorted {\kago} geometry
that would be identified in the future)
and thus remain worth consideration.
In order to get insight into the influence of the anisotropy
on the nature of the gound state and the low energy excitations,
exact diagonalization have been carried out for clusters up to $N=36$
spins which will be reported in an other paper~\cite{ps_akag}.
Preliminary investigations of the modification to the Heisenberg model
that are required for a better description of Volborthite
have been also started and will be be persued.
First principle investigations might help to 
provide further inside into this question and 
more experimental data would be very usefull.

\section*{Acknowledgements}
I am grateful to Zenji Hiroi for providing its data
and to W.~Appel, F.~Bert, P.~Mendels, G.~Misguich,
H.-U. Everts, T. Yavors'kii for discussions
on the anisotropic {\kago} model or Volborthite.

\end{document}